\documentclass[10pt]{article}
\usepackage{graphicx}
\usepackage{hyperref}
\usepackage{amsmath}
\usepackage{textcomp}

\begin{document}

\LARGE
\bfseries
Triple resonant four-wave mixing boosts the yield of continuous coherent VUV generation

\normalsize

\vspace{1cm}
\begin{bfseries}
D. Kolbe, M. Scheid$^*$, and J. Walz
\end{bfseries}

\vspace{1cm}
\itshape
Institut f{\"{u}}r Physik, Johannes Gutenberg-Universit\"{a}t Mainz and Helmholtz Institut Mainz, D-55099 Mainz, Germany

$^*$Present address: Carl Zeiss Laser Optics GmbH, Carl Zeiss Strasse, D-73446 Oberkochen, Germany 

\vspace{1cm}
\upshape
   
Continuous-wave coherent radiation in the vacuum ultraviolet (VUV) wavelength region at 121\,nm will be essential for future laser-cooling  of trapped antihydrogen \cite{Andresen10}. Cold antihydrogen will enable both tests of the fundamental symmetry between matter and antimatter at unprecedented experimental precision \cite{Haensch93} and also experiments in antimatter gravity \cite{Perez05}. Another fascinating application of narrowband continuous laser radiation in the VUV is quantum information processing using single trapped ions in Rydberg-states \cite{Zoller08,SchmidtKaler11}. Here we describe highly efficient continuous four-wave mixing in the VUV by using three different fundamental wavelengths with a sophisticated choice of detunings to resonances of the nonlinear medium. Up to 6 microwatts of vacuum ultraviolet radiation at 121\,nm can be generated which corresponds to an increase of three orders of magnitude in efficiency.

\vspace{1cm}

\mdseries

VUV laser radiation can be generated by four-wave sum-frequency mixing (FWM) in gases and metal vapours \cite{Wallenstein80,Marangos90,Vidal92}, a process in which three laser fields generate a nonlinear polarisation which acts as the driving term for the fourth coherent field at the sum-frequency. A two-photon resonance is kown to dramatically enhance the FWM efficiency \cite{Smith88}. To further increase the efficiency the use of one-photon resonances is desirable. However, the dispersion of close one-photon resonances affects phasematching between the driving nonlinear polarization and the generated field in the VUV which can severely limit the efficiency of the FWM process. Here, we show that the interplay of two resonances at one- and three-photon height can be used to cancel phasematching limitations and to take full benefit of the enhancement in the nonlinear susceptibility. Powers of up to 6\,\textmu W at 121.26\,nm wavelength were achieved. This is 30 times more power than previously reported \cite{Pahl05} and three orders of magnitude more efficient.

\begin{figure*}[htb]
\centerline{\includegraphics[width=13.5cm]{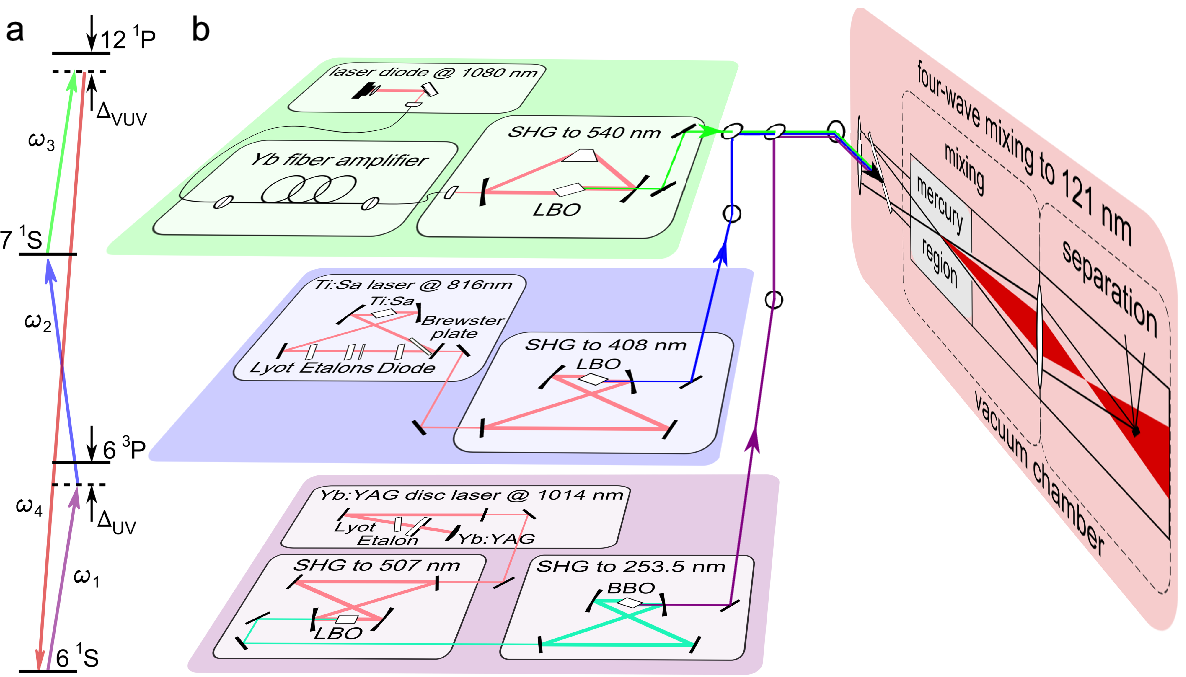}}
\caption[Level scheme] {\textbf{Energy-level diagram of mercury, the FWM scheme and setup.} \textbf{a}, The UV laser (254\,nm) is tuned close to the $6^1S$--$6^3P$ resonance and the blue laser beam (408\,nm) establishes the two photon resonance with the $7^1S$ state. The green laser is at 540\,nm, so that the resulting wavelength is close to the $6^1S$--$12^1P$ resonance. \textbf{b}, The fundamental beams at 254\,nm, 408\,nm and 540\,nm wavelength are produced by frequency-doubling and frequency-quadrupling of strong continuous-wave solid-state infrared lasers. The beams are shaped, overlaped and focused into the mercury cell. In the FWM region VUV generation takes place. The VUV beam is separated from the fundamental beams by the dispersion of a MgF$_2$ lens and four VUV filters. A solar-blind photomultiplier tube (not shown) is used for detection. (SHG: \emph{second harmonic generation}, LBO: \emph{lithium triborate crystal}, BBO: \emph{$\beta$-barium borate crystal})}\label{Fig:1}
\end{figure*}

Our experiment uses mercury vapour as a nonlinear medium and the relevant energy levels are shown in Fig. \ref{Fig:1}(a). A UV beam at 254\,nm and a blue beam at 408\,nm wavelength are tuned to the two-photon resonance between the 6$^1S$ ground state and the 7$^1S$ state. The third beam at 540\,nm can be tuned such that the sum frequency of the three fundamental beams is near the 6$^1S$--12$^1P$ resonance, which is at 121.26\,nm wavelength. 

A schematic of the experimental setup for the VUV production is shown in Fig.~\ref{Fig:1}(b). The three fundamental beams are overlaped using dichroic mirrors and focused into the mercury cell which provides a mercury atom density of up to $N=1.1\times 10^{24}\textrm{m}^{-3}$. Outside the focus region cooled baffles and helium buffer gas with a pressure of 20\,mbar are used to avoid condensation of mercury on the optics.


The power at the sum frequency $\omega_4=\omega_1+\omega_2+\omega_3$ is given by \cite{Bjorklund75}:

\begin{equation}\label{FWMeqn}
P_4=\frac{9}{4}\frac{\omega_1\omega_2\omega_3\omega_4}{\pi^2\epsilon^2_0 c^6}\frac{1}{b^2}\left(\frac{1}{\Delta k_a}\right)^2 \left|\chi^{(3)}_a\right|^2 P_1 P_2 P_3 G(b N \Delta k_a).
\end{equation}

Here $\omega_i$ is the angular frequency, $P_i$ the power of the $i$th beam, $N$ the density of the nonlinear medium, $\chi^{(3)}_a$ the nonlinear susceptibility per atom and $\Delta k_a=(k_4-k_1-k_2-k_3)/N$ the wavevector mismatch per atom. $G(b N \Delta k_a)$ is the phasematching function for phasematching by the density of the nonlinear medium with a maximum value at $b N \Delta k_a =-4$ \cite{Bjorklund75}. Equation (\ref{FWMeqn}) describes a four-wave mixing process in a non-absorbing gaseous medium with tight focused fundamental beams with equal confocal parameters $b$.

Two important factors are the nonlinear susceptibility $\chi^{(3)}_a$ and the wavevector mismatch $\Delta k_a$, which are both independent of the density but functions of the fundamental and sum frequencies. The leading term in the nonlinear susceptibility per atom is \cite{Smith87}:

\begin{equation}
\chi^{(3)}_a=\frac{1}{6 \epsilon_0 \hbar^3} \sum_{m,n,v}{ \frac{1}{\omega_{ng}-(\omega_1+\omega_2)}\frac{\rho_{nm}\rho_{mg}}{\omega_{gm}-\omega_1} \frac{\rho_{nv}\rho_{vg}}{\omega_{gv}-\omega_4}}.
\label{chi}
\end{equation} 

\begin{figure}[htb]
\centerline{\includegraphics[width=8.5cm,angle=-90]{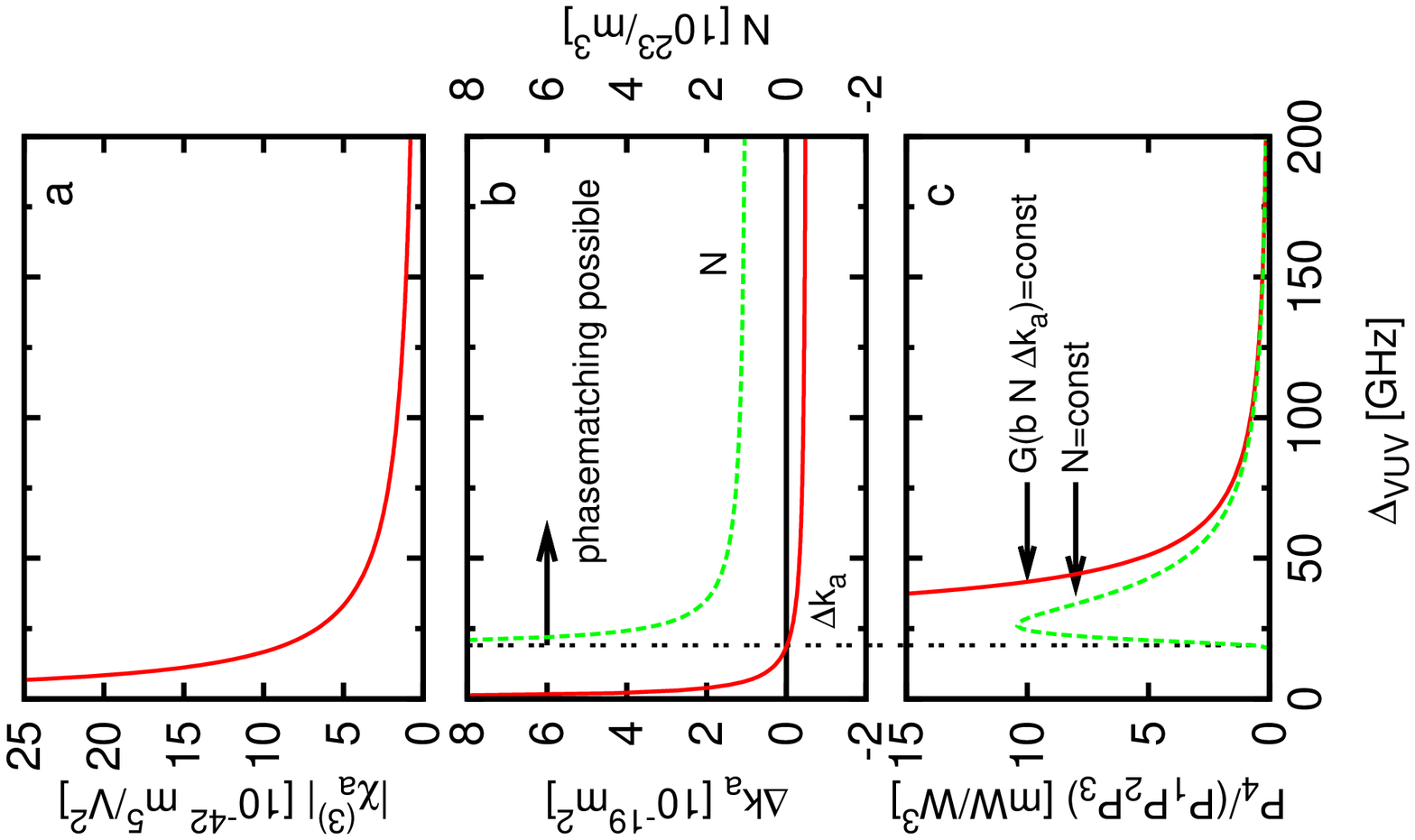}}
\caption{\textbf{Calculation of the FWM power.} \textbf{a}, nonlinear susceptibility, \textbf{b}, wavevector mismatch (red solid) and phasematching density (green dashed), \textbf{c}, FWM efficiency at perfect phasematching (red solid) and constant density (green dashed) versus the detuning of the FWM frequency. The UV detuning to the $6^3P$ level is 200\,GHz.}\label{Fig:2}
\end{figure}

Here $\omega_{ij}$ is the transition frequency from the initial state $i$ to the final state $j$ and $\rho_{ij}$ is the corresponding dipole matrix moment. The states $m$ and $v$ are linked to the ground state $g$ via a dipole transition at $\omega_1$ and $\omega_4$, respectively. State $n$ is connected to the ground state via a two-photon transition. In the case of FWM with a two-photon resonance and near one-photon resonances for both one fundamental and the generated beam, most of the terms in the sum are negligible and the nonlinear susceptibility is proportional to products of terms related to both resonances. Divergencies at resonances are damped by radiative deacay etc., which corresponds to the linewidths. This is not shown in equation (\ref{chi}) but taken into account in the calculations. The strong enhancement of the nonlinear susceptibility $\chi^{(3)}_a$ near the $6^1S$--$12^1P$ resonance is illustrated in Fig.~\ref{Fig:2}(a).
 
The wavevector mismatch is dominated by the UV and VUV beam because they are close to one-photon resonances and can be described by the real part of the linear susceptibilities  $\chi^{(1)}$:

\begin{equation}
\begin{split}
\Delta k_a &=\frac{1}{2 N c} \left(\omega_4 \Re \left\{\chi^{(1)}(\omega_4)\right\} - \omega_1 \Re\left\{\chi^{(1)}(\omega_1)\right\}\right)\\
          &=\frac{1}{3 c \epsilon_0 \hbar} \sum_v{ \omega_{gv} \rho_{gv}^2 \left(\frac{\omega_4}{\omega_{gv}^2-\omega_4^2}-\frac{\omega_1}{\omega_{gv}^2-\omega_1^2}\right)}.
\end{split}
\label{wavevector}
\end{equation}

Here the contribution of the blue and green beams are neglected and $v$ represents all the states which are connected to the ground state via a dipole transition. In the near one-photon resonant approximation just the states near a one-photon resonance are kept in the sum over $v$. When both frequencies are red-detuned to resonances both terms are positive. To achieve a negative $\Delta k_a$ (which is necessary for phasematching) the second term, which is due to the resonance of the fundamental beam, has to be larger than the first term, which is due to the resonance of the generated FWM beam. In Fig. \ref{Fig:2}(b) the wavevector mismatch $\Delta k_a$ is shown as a function of the VUV detuning (red solid curve). Near the one-photon resonance it changes sign and phasematching can not be achieved by adjusting the density of the nonlinear medium. This can be also seen by the divergence of the phasematching density (green dashed line in Fig. \ref{Fig:2}(b)) at the zero crossing of $\Delta k_a$.

The power at the FWM frequency is proportional to $\left( |\chi^{(3)}_a| / \Delta k_a \right)^2$ when the phasematching function $G(N b \Delta k_a)$ is maximized and is shown in Fig. \ref{Fig:2}\,(c). The red solid line represents the power at perfect phasematching. Here the density has to be adjusted as shown in (b). However, in the experiment a too strong increase of the mercury density is not useful because of absorption. For the green dashed line we took this into account and fixed the density at the phasematching density in the case of FWM far from the one-photon resonance ($N=1\times10^{23}$m$^{-3}$). The strong increase in the FWM efficiency is due to the combination of two single one-photon resonances: one (UV) to dominate the phasematching and one (VUV) to take full benefit of the increase in the nonlinear susceptibility.


Experimentally there are three free parameters for the FWM process: The detuning of the UV beam to the $6^1S$--$6^3P$ transition ($\Delta_{\textrm{UV}}$), the detuning of the VUV beam to the $6^1S$--$12^1P$ transition ($\Delta_{\textrm{VUV}}$), and the density of the mercury vapour ($N$), which can be adjusted by changing the temperature of the mercury cell. The density affects both the phasematching of the FWM process and the absorption of the VUV beam. The detunings affect the nonlinear susceptibility and the wavevector mismatch and thus also the phasematching. The optimal density for FWM at a specific set of detunings results from a combination of nearly perfect phasematching and small absorption. To separate these two effects we have varied the position of the focus in the mercury vapour. This changes the path length of the VUV beam in the mercury vapor region and thus the absorption length. The absorption coefficient at the VUV wavelength is obtained from the exponential decrease in the FWM efficiency. 

\begin{figure}[htb]
\centerline{\includegraphics[width=8.5cm,angle=-90]{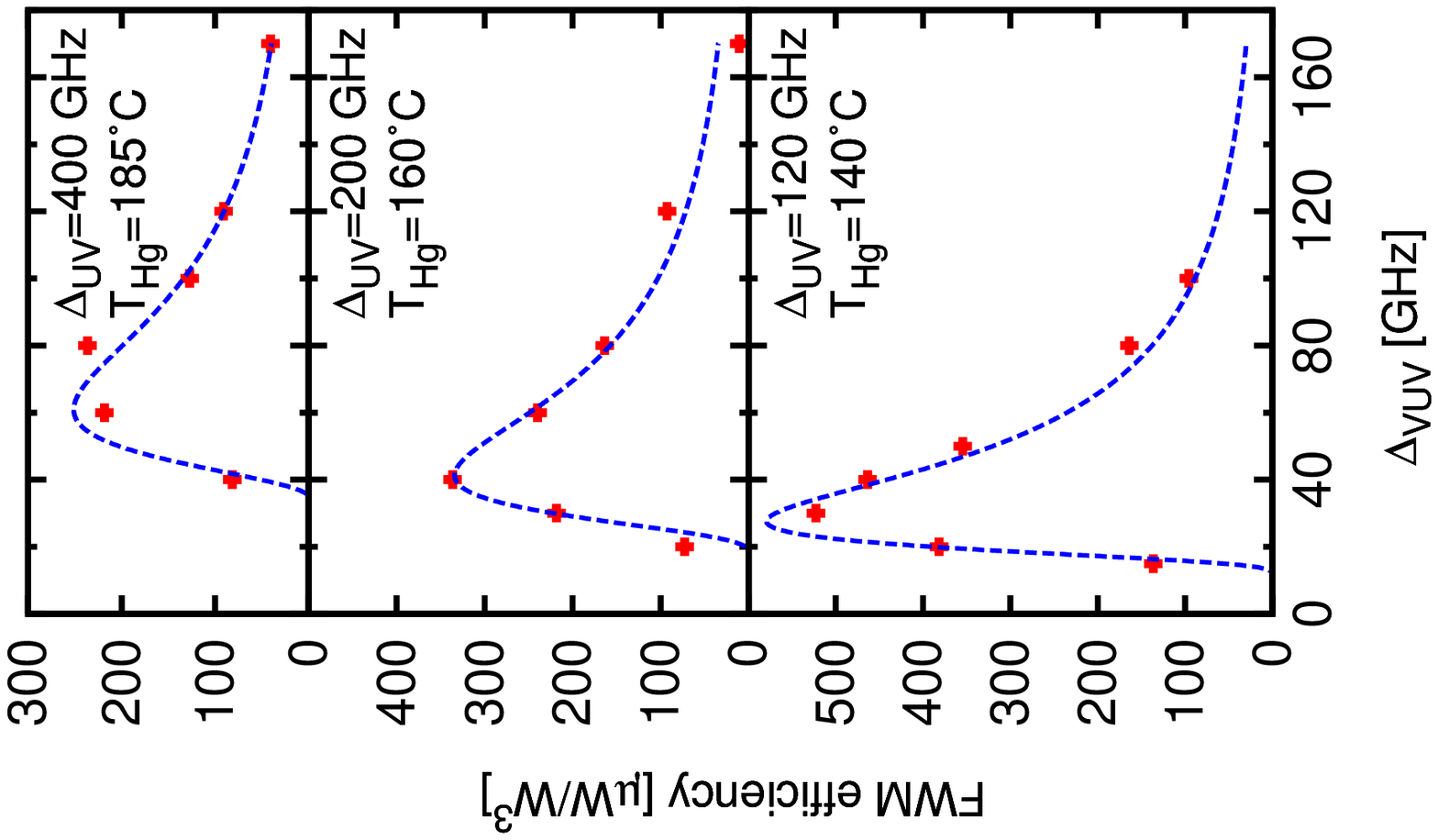}}
\caption{\textbf{Triple resonant FWM.} FWM efficiency as a function of the detuning to the $12^1P$ state at different UV detunings to the $6^3P$ state. The dashed lines are calculations which include imperfect beam overlap of the fundamental beams, absorption at the VUV frequency and saturation.}
\label{Fig:4}
\end{figure}

The nonlinear susceptibility takes full benefit of smaller detunings at the VUV wavelength, if the dispersion at the VUV wavelength does not adversely effect the phasematching, i.\,e. the wavevector mismatch in Eq.~\ref{wavevector} is still negative because of the dispersion at the UV wavelength (second term). Smaller UV detunings enable a larger FWM efficiency due to the possibility to choose smaller VUV detunings since they contribute with opposite signs to the wavevector mismatch. This is observed in the experiment and shown in Fig.~\ref{Fig:4}. At every UV detuning the FWM efficiency gains by choosing smaller VUV detunings. The maximum in the FWM efficiency depends on the UV detuning and changes from $\Delta_{\textrm{VUV}}=60$\,GHz at $\Delta_{\textrm{UV}}=400$\,GHz to $\Delta_{\textrm{VUV}}=30$\,GHz at $\Delta_{\textrm{UV}}=120$\,GHz. At detunings smaller than the optimum the VUV yield decreases for two reasons: Absorption of the VUV beam due to the near resonance and a change in the phasematching.

The dashed lines in Fig. \ref{Fig:4} are calculations of the FWM efficiency which include absorption at the VUV and UV wavelength. For every combination of detunings the nonlinear susceptibility per atom $\chi^{(3)}_a$ and the wavevector mismatch per atom $\Delta k_a$ are calculated. The FWM equation (\ref{FWMeqn}) does not include absorption. Absorption of the fundamental beams is taken into account by using the fundamental powers at the position of the focus and thus at the FWM region. Absorption of the VUV beam is included by an exponential decay from the focus on over the length of the mercury cell. The influence of absorption on the FWM phasematching can be neglected if absorption within the Rayleigh region of the fundamental beams is small. At a UV detuning of 50\,GHz and a density of $3.8 \times 10^{22}/$m$^{-3}$ we get about 3\% of absorption within the Rayleigh length of $0.8$\,mm. 

Let us now compare the absolute value of the observed efficiency in the experiment at a UV detuning of 200\,GHz, Fig.~\ref{Fig:4} center, with the calculation for an ideal situation and without absorption, Fig.~\ref{Fig:2}. The reduction by a factor of about 17 is due to three effects: (1.) Imperfect overlap of the fundamental beams and fractional power in the lowest Gaussian beam mode \cite{Kolbe11} causes an overall reduction in the efficiency by a factor  of 4. (2.) Absorption in the VUV. The absorption length is 5.1\,mm at a VUV detuning of 40\,GHz, which has been measured by changing the focus position in the mercury vapour. This corresponds to a reduction of 3 in the VUV yield. (3.) Saturation has been observed by changing the fundamental powers and measuring the change in FWM efficiency. (No such saturation was observed in earlier experiments \cite{Kolbe11,Scheid09} because they were at much larger VUV detunings.) Saturation contributes a factor of 1.4. A maximum yield of 6\,\textmu W in the VUV was achieved with input powers of 182\,mW(UV), 245\,mW(blue), 525\,mW(green) and detunings of $\Delta_{\textrm{UV}}=120$\,GHz and $\Delta_{\textrm{VUV}}=40$\,GHz.

In conclusion, we have demonstrated a high power continuous-wave laser source at 121\,nm wavelength. Utilizing the interplay of two resonances at one- and three-photon height a large increase in FWM efficiency could be achieved. We observed powers up to 6\,\textmu W, which is 30 times more than ever achieved in this wavelength region and three orders of magnitude more efficient.

\section*{Methods}
\begin{small}

\begin{bfseries}
Lasersystem for the fundamental beams.
\end{bfseries}
A schematic of the laser system to produce the three fundamental beams is shown in the lower part of Fig. \ref{Fig:1}(b). The beam at $254\,\text{nm}$ is produced by a frequency-quadrupled Yb:YAG disc laser (ELS, VersaDisk 1030-50). Frequency-quadrupling is done with two resonant enhancement cavities, the first one using a lithium triborate crystal (LBO) as nonlinear medium, the second one using a $\beta$-barium borate crystal (BBO). From $2\,\text{W}$ of infrared light at $1015\,\text{nm}$ we get up to $200\,\text{mW}$ of UV radiation. This system is capable of producing up to 750\,mW of UV light, for details see \cite{Scheid07}. The second fundamental beam at $408\,\text{nm}$ is produced by a frequency-doubled titanium:sapphire laser (Coherent, 899-21), pumped by a frequency doubled Nd:YVO$_4$ laser (Coherent, V10). The external frequency-doubling cavity uses a LBO crystal. From $1.5\,\text{W}$ of infrared (IR) light at $816\,\text{nm}$ we get up to $500\,\text{mW}$ of blue light. The third fundamental beam at $540\,\text{nm}$ is produced with a grating stabilized diode laser at 1080\,nm boosted by a fiber amplifier system (Koheras, Boostik) and frequency doubled by a modified commercial frequency-doubling cavity (Spectra Physics, Wavetrain). This system is capable of producing up to $4\,\text{W}$ of green light at 545.5\,nm \cite{Markert07}. For the present experiments we operate the IR laser at $740\,\text{mW}$, which gives $280\,\text{mW}$ of green light at 540\,nm. 

\begin{bfseries}
Wavelength separation \& detection.
\end{bfseries}
The four-wave mixing region is separated from the detection region by a vacuum sealed MgF$_2$ lens which also performs the wavelength separation of the VUV light from the fundamental beams (see Fig. \ref{Fig:1}(b)). Due to the dispersion of this lens the focal length differs for the VUV wavelength ($f=21.5$\,cm at 540\,nm, $f=13$\,cm at 121\,nm). A tiny mirror is placed in the focus of the fundamental beams to reflect them to the side. The VUV beam is large at the fundamental focus ($w \approx 4.9$\,mm) and therefore the mirror just casts a shadow in the VUV beam, causing $\approx 30\%$ loss. A solar-blind photomultiplier tube is used for detection of the VUV photons. The background is suppressed by four 121\,nm filters. The overall detection efficiency due to losses in the MgF$_2$ lens, the tiny mirror, the four filters and the photomultiplier efficiency is $9 \times 10^{-7}$.  

\end{small}

\section*{Acknowledgments}
The authors acknowledge support for this work from the Deutsche Forschungsgemeinschaft and the Bundesministerium f\"{u}r Bildung und Forschung.

\section*{Author contributions}
All authors contributed significantly to the work presented in this paper.

\section*{Additional information}
The authors declare no competing financial interests.

\end{document}